\documentclass[reprint,aps,pra,superscriptaddress,twocolumn]{revtex4-1}

\usepackage{amsmath}
\usepackage{graphicx}
\usepackage{hyperref}

\begin{document}

%\preprint{}

\title{Determination of any pure spatial qudits from a minimum number of measurements by phase stepping interferometry.}



\author{Quimey Pears Stefano}
\email[]{quimeyps@df.uba.ar}
\affiliation{Universidad de Buenos Aires, Facultad de Ciencias Exactas y Naturales, Departamento de F\'isica, Buenos Aires, Argentina}
\affiliation{Consejo Nacional de Investigaciones Cient\'ificas y T\'ecnicas, Buenos Aires, Argentina.}

\author{Lorena Reb\'on}
\affiliation{Departamento de F\'isica, IFLP-CONICET, Universidad Nacional de La Plata, C.C. 67, 1900 La Plata, Argentina.}

\author{Silvia Ledesma}
\affiliation{Universidad de Buenos Aires, Facultad de Ciencias Exactas y Naturales, Departamento de F\'isica, Buenos Aires, Argentina}
\affiliation{Consejo Nacional de Investigaciones Cient\'ificas y T\'ecnicas, Buenos Aires, Argentina.}

\author{Claudio Iemmi}
\affiliation{Universidad de Buenos Aires, Facultad de Ciencias Exactas y Naturales, Departamento de F\'isica, Buenos Aires, Argentina}
\affiliation{Consejo Nacional de Investigaciones Cient\'ificas y T\'ecnicas, Buenos Aires, Argentina.}

%\thanks{}


\date{\today}

\begin{abstract}
We present a proof-of-principle demonstration of a method to
characterize \textit{any} pure spatial qudit of arbitrary dimension
$d$, which is based on the classic phase shift interferometry
technique. In the proposed scheme a total of only $4 d$ measurement
outcomes are needed, implying a significant reduction with respect to
the standard schemes for quantum state tomography which require of the
order of $d^2$. By using this technique, we have experimentally
reconstructed a large number of states ranging from $d=2$ up to $14$
with mean fidelity values higher than $0.97$. For that purpose the
qudits were codified in the discretized transverse momentum-position
of single photons, once they are sent through an aperture with $d$
slits. We provide an experimental implementation of the method based
on a Mach-Zehnder interferometer, which allows to reduce the number of
measurement settings to 4 since the $d$ slits can be measured
simultaneously. Furthermore, it can be adapted to consider the
reconstruction of the unknown state from the outcome frequencies of
$4d-3$ fixed projectors independently of the encoding or the nature of
the quantum system, allowing to implement the reconstruction method in
a general experiment.

\end{abstract}

 % Optical implementations of quantum information
                % processing and transfer
 % Optical processors, correlators, and modulators
 % State reconstruction, quantum tomography
 % Phase shifting interferometry

% \pacs{42.50.-p} % Quantum optics

% \pacs{42.50.Dv} % Quantum state engineering and measurements (see also
%                 % 03.65.Ud Entanglement and quantum nonlocality, e.g.,
%                 % EPR paradox, Bells inequalities, GHZ states, etc.)
% \pacs{42.79.Kr} % Display devices, liquid-crystal devices (see also
%                 % 85.60.Pg Display systems)
% \pacs{03.65.Wj} % State reconstruction, quantum tomography

% \pacs{03.67.-a} % Quantum information (see also 42.50.Dv Quantum state
%                 % engineering and measurements; 42.50.Ex Optical
%                 % implementations of quantum information processing and
%                 % transfer in quantum optics)

%\keywords{}

\maketitle
\newcommand{\OBJ}{$\mathbf{OBJ}$}

\newcommand{\SFcero}{$\mathbf{SF0}$}
\newcommand{\SFuno}{$\mathbf{SF1}$}
\newcommand{\SFdos}{$\mathbf{SF2}$}

\newcommand{\BSuno}{$\mathbf{BS1}$}

\newcommand{\Luno}{$\mathbf{L_1}$}
\newcommand{\Ldos}{$\mathbf{L_2}$}
\newcommand{\Ltres}{$\mathbf{L_3}$}

\newcommand{\Limg}{$\mathbf{L_\mathrm{img}}$}
\newcommand{\Lft}{$\mathbf{L_\mathrm{ft}}$}

\newcommand{\ImagePlaneA}{$\pi'$}
\newcommand{\ImagePlaneB}{$\pi''$}
\newcommand{\ImagePlaneC}{$\pi'''$}

\newcommand{\IA}{$\mathbf{IA}$}
\newcommand{\FA}{$\mathbf{FA}$}

\newcommand{\SLM}{$\mathbf{SLM}$}

\newcommand{\unitMuM}[1]{#1\,\mathrm{\mu m}}
\newcommand{\unitCM}[1]{#1\,\mathrm{cm}}

\newcommand{\figref}[1]{Figure \ref{#1}}

\newcommand{\ket}[1]{\ensuremath{|#1\rangle}}
\newcommand{\bra}[1]{\ensuremath{\langle #1|}}
\newcommand{\braket}[2]{\ensuremath{\langle#1|#2\rangle}}
\newcommand{\ketbra}[2]{\ensuremath{|#1\rangle\!\langle#2|}}

\newcommand{\Projector}[1]{\ensuremath{\mathrm{Proj}\left(#1\right)}}

\newcommand{\PP}[2]{\ensuremath{\mathrm{\hat{P}}_{#1}^{(#2)}}}

\section{Introduction}
Determining the state of a quantum system is one of the fundamental
tasks in quantum information processing and a recurrent problem in
quantum mechanics \cite{Paris2010}. In this regard, quantum state
tomography provides a mean of fully reconstructing the density matrix
which describes the state of a quantum system. For typical quantum
state tomography methods
\cite{James2001,Thew2002,Wootters1989,Adamson2010} the number of
required measurements settings (or outcomes) increases with the
dimension of the system $d$, as $d^2$, that makes difficult the
treatment of high-dimensional quantum systems. Therefore, as diverse
applications of quantum information can be enhanced using a dimension
greater than two
\cite{Cerf2002,Collins2002,Groblacher2006,Dixon2012,Lee2016}, there is
a growing interest in estimating $d$-level quantum systems (qudits)
from a reduced number of measurements.

With some \emph{a priori} information of the unknown quantum system, a
reduction in the number of measurements is feasible.  For example,
in case of pure or nearly pure quantum states, compressed sensing
techniques allow obtaining, with high probability, the reconstruction
of the state with a number of measurements of the order of
$d(\log d)^2$ \cite{Gross2010,Gross2016}. This technique works by
randomly choosing a set of observables and measuring their expectation
values. Thus, it does not provide an explicit measurement
set-up. Besides, the amount of measurements is still far to be
optimal.

Flammia \emph{et al.}\ \cite{flammia2005minimal} had established that a
measurement with at least $2d$ outcomes is required to determine
\textit{almost all} (but not all) pure states. Furthermore, they have
also demonstrated that $3d$-$2$ one-dimensional projectors are
sufficient for determining a generic pure state, with the exception of
a set of measure zero. This number increases if we want to distinguish
\textit{any} two pure states. In such a case, a measurement with
$\sim4d$ outcomes must be considered \cite{Heinosaari2013}, or when
restricting to projective measurements, at least four orthonormal
bases are required if $d\geq3$, except maybe for $d=4$, in which case
it is not known whether three bases would be sufficient
\cite{Carmeli2015}. However, the measurements do not provide a way of
verifying the purity assumption.

Recently, Goyeneche \emph{et al.} \cite{Goyeneche2015} have proposed a
method to determine an arbitrary pure state of any dimension by means
of projective measurements onto five fixed orthonormal bases,
resulting in a total of $5d$ measurements outcomes. They have
experimentally implemented the method for reconstructing spatial
qudits \cite{Neves2004}. The measurement settings required for that
scheme could be interpreted as equivalent to a \textit{four step}
phase shifting interferometry (PSI) between pairs of consecutive
slits. As it is well known PSI leads to the most accurate way to
measure the amplitude and phase distribution of a wavefront
\cite{Creath1988}. In these techniques controlled phase displacements
are introduced between the reference and the object beam, then the
wavefront under test can be determined from the interferograms
corresponding to the different phase shifts. The number of
interferograms to be recorded, as the phase is shifted, varies
depending on the algorithm employed to recover the phase distribution
of the wavefront. Typically, four or three step algorithms are
used.

In dimension $d=2$, the connection between quantum state tomography
and PSI was studied by Reb\'on \emph{et al.} \cite{Rebon2013}. They
showed that for this particular case the full quantum tomography of
any arbitrary qubit, pure or mixed, is equivalent to a four step
PSI. In that work, a path-qubit was codified as the superposition
state of a single-photon occupying two arms of a Michelson
interferometer. The PSI was carry out by obtaining the different
interferograms between both paths with one of them as the reference.

In this article we propose a quantum state estimation method, based in
a \textit{three step} PSI algorithm, that allows to determine any pure
spatial qudit of arbitrary dimension $d$ by means of a minimum number
of measurements. In fact, in our method the number of measurement
bases is 4, which is lower than the number of bases requiered in
Ref. \cite{Goyeneche2015}, and even more, it is consistent with the
minimum number of measurement outcomes reported in
\cite{flammia2005minimal,Heinosaari2013,Carmeli2015}. In Section
\ref{method} we provide a complete description of the qudit estimation
process and point out how the measurements onto these 4 basis are also
sufficient for verifying the purity assumption of the unknown
state. In addition, for photonic qudits codified in the transverse
momentum-position of single photons, we provide an experimental
implementation of the method based on an interferometer scheme. Our
setup allows to reduce the number of measurement settings to only 4,
regardless of the dimension $d$ of the system. The results are
presented and discussed in Section \ref{results}, before
going into the conclusions.

\section{Method}\label{method}
The encoding process of the $d$-dimensional quantum system is
performed in the discretized transverse momentum of single photons
once they are sent through an aperture with $d$ slits
\cite{Neves2005,Lima2009}. Such pure state can be expressed as
\begin{eqnarray}\label{qudit}
|\Psi\rangle =\sum_{k=0}^{d-1} c_k |k\rangle,
\end{eqnarray}
where the $c_{k}~'s$ are the complex coefficients that represent the
complex transmission amplitude of each slit, and $|k\rangle$ denotes
the state of the photon passing through the slit $k$. These
coefficients can be explicitly written as $c_k=|c_k| e^{i \varphi_k}$
where $\varphi_k$ represents the argument of the complex number
$c_k$. For reconstructing the quantum state of these systems we use
one of the slits as a phase reference and implement the three step PSI
algorithm to find the phase of each of the remaining slits with
respect to the reference, that is finding the argument $\varphi_k$.
The additional measurement of the intensity of each slit allows the
unambiguous reconstruction of the state up to an arbitrary global
phase and also give us a way to certify if the state is pure -- or
nearly pure-- without any a priori assumptions. The total number of
measurements outcomes in this method is $4d-3$ when the procedure is
performed in an adaptive way, or $4d$ in case of fixed measurement
settings. Even more, the proposed experimental setup for
reconstructing spatial qudits has the advantage that every of the four
sets of $d$ measurements corresponds to a single interferogram, thus,
using photon counting cameras \cite{Krishnaswami2014} instead of a
point-like single photon detection module (SPDM), $d$ measurements can
be recorded in only one acquisition, i.e. only 4 pictures are needed,
in any dimension $d$, to determine the unknown state. Nevertheless,
the set of $4d-3$ quantum projectors to be used in order to perform
the tomographic process do not depend on the particular encoding or
the nature of the quantum system and they could be applied in a
completely general setup.

\begin{figure}
\includegraphics{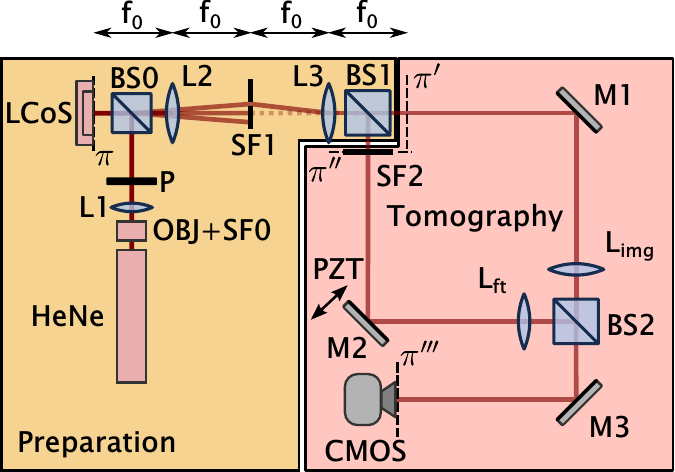}
\caption{Experimental setup for reconstructing pure spatial
  qudits. \textit{Preparation part}: an expanded and collimated HeNe
  laser impinges onto a \emph{phase only} LCoS modulator. In
  conjunction with the $4f$ processor formed by the lens \Ldos\ and
  \Ltres\, and the spatial filter \SFuno\, the quantum state is
  encoded in the planes \ImagePlaneA\ and \ImagePlaneB\
  . \textit{Tomography part:} A lens \Limg\ onto the \emph{Image arm}
  of the Mach Zender interferometer images the plane \ImagePlaneA\
  onto the output plane \ImagePlaneC\ . The lens in the \emph{Fourier
    arm} \Lft\ perform the Fourier Transform of the only slit that is
  not blocked by the spatial filter \SFdos\
  . \label{fig:experimental-setup}}
\end{figure}

Let us to start by briefly describing the state preparation which is
carried on by using the first part of the optical setup sketched in
Fig. \ref{fig:experimental-setup}. The light source is an HeNe laser
that is expanded, filtered, and collimated by the objective \OBJ, the
spatial filter \SFcero\ and the lens \Luno. To test the proposed
method at the single-photon level we inserted neutral-density filters
to highly attenuate the power of the laser beam to 0.005 nW.  It
implies that, for an interferometer with a total length of 140 cm as
in our case, less than one photon on average is present, at any time,
in the experiment. This source can be used to mimic the single-photon
qudit state given by Eq.~(\ref{qudit}), and as is usual in optical
implementations of quantum-states estimation, it is enough to test the
feasibility of the proposed method
\cite{Leach2002,Lima2011,Malik2014}.  The beam that impinges on the
spatial light modulator used to codify the slit states, has
approximately constant amplitude and phase over the regions of
interest (ROIs) where the slits are displayed.  The method for
codifying \emph{arbitrary} complex amplitudes of spatial photonic
qudits was developed for our group in previous
works \cite{Solis-Prosser2013,Varga2014}.  We briefly explain here the
main features of the method: Blazed phase gratings
are displayed onto each slit region. The real amplitude of the slit is
determined by the diffraction efficiency achieved through the phase
modulation of the grating. On the other hand the desired phase value
is obtained just by adding an adequate constant phase. The required
pure phase modulation is provided by a parallel aligned liquid crystal
on silicon display (LCoS) Holoeye PLUTO with HDTV resolution
(1920x1080) and pixel size of $\unitMuM{8}.$ In our case the width of
the slits is 10 pixels, and the separation between slit centers is 30
pixels. In order to implement the mentioned codification we use a
typical $4f$ processor conformed by lenses \Ldos\ and \Ltres\
($f_0 = \unitCM{20}$).  The spatial period of the gratings displayed
onto the slit regions is 16 pixels which is enough to select by means
of the spatial filter \SFuno\ the first diffracted order. This optical
setup together with the non polarizing beam splitter \BSuno\ allows to
obtain on planes \ImagePlaneA\ and \ImagePlaneB\ the desired complex
amplitude distribution.

The tomographic process employed to characterize the $d$-dimensional
spatial qudit is implemented by using the Mach-Zehnder interferometer
schematized in the second part of Fig.
\ref{fig:experimental-setup}. Let us call \emph{Image Arm} (\IA) the
one that contains lens \Limg. This lens in configuration $2f - 2f$
($f_\mathrm{img}= \unitCM{35}$) images the input state obtained on
\ImagePlaneA\ over the final plane \ImagePlaneC. Meanwhile the
\emph{Fourier Arm} (\FA) is the one that contains the lens \Lft\ in
configuration $f - f$ ($f_\mathrm{ft} = \unitCM{70}$) giving the exact
\emph{Fourier Transform} of plane \ImagePlaneB\ over \ImagePlaneC. The
spatial filter \SFdos~(a slit of width $\unitMuM{200}$), placed on
plane \ImagePlaneB, blocks all but one slit that acts as reference.
The resulting output is the interference pattern between the complex
amplitude of the $d$ slits and the reference.  Finally, intensity
measurements are carried out by means of a high sensitive camera based
on \textbf{CMOS} technology placed in \ImagePlaneC. The camera used is
an Andor Zyla 4.2 sCMOS.

\begin{figure}[htbp]
%\centering
  \includegraphics[width=\linewidth]{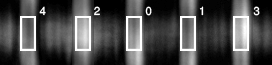}\caption{Interferogram
    for a state of dimension $d=5$. The vertical lighted bands
    correspond to the image of the five slits. The horizontal lighted
    band corresponds to the Fourier Transform of the filtered slit in
    the FT arm of the interferometer, which acts as the reference. The
    rectangles indicate the regions on which the measurements are
    performed, and $0, 1,...,4$ are the corresponding slit
    numbers.\label{fig:exp_photo_interf}}
\end{figure}

As an example, one of the interferograms obtained at the output of the
Mach-Zender for a qudit of dimension $d=5$ is shown in Fig.
\ref{fig:exp_photo_interf}. The lighted bands in the vertical
direction correspond to the image of the five slits. The lighted band
in the horizontal direction is the Fourier Transform of the only slit
that is not blocked by \SFdos\ in the \FA\ of the interferometer. This
is the slit that acts as the reference. The drawn rectangles delimit
the ROIs on which the measurements are performed.

It is important to note that a similar implementation of the
reconstructing method can be done with exactly the same setup by using
a SPDM which must be displaced over the final plane \ImagePlaneC\ in
order to measure \textit{sequentially} the counts in the different
ROIs. However, the use of high sensitive cameras, that increasingly
become an interesting option for single photon detection in quantum
optics experiments \cite{Leach11,fickler2013real,Unternahrer16}, makes
possible to complete the measurement stage by taking 4 snapshot, no
matter the dimensionality of the unknown spatial qudit. This is
possible both due to the proposed setup which enables to perform a
\textit{simultaneous} detection of the $d$ regions (see
Fig.~\ref{fig:exp_photo_interf}) as well as the selected PSI scheme.
In fact, a simultaneous measurement is not possible by using the set
of measurement bases presented in \cite{Goyeneche2015} since in such a
case, the tomographic process is equivalent to a PSI scheme which
requires the sequential interference of contiguous slits, i. e., there
is not an unique reference beam as in our case.

We will now proceed to analyze the tomographic reconstruction
method. In order to characterize the quantum state in
Eq.~(\ref{qudit}) it is necessary to know the complex amplitudes
$c_k$, i.e., the amplitude and phase of the wavefront just in the
region the slit $k$. To this end we implemented the classical PSI
technique of three steps, involving successive phase shifts of $\pi/2$
that were introduced in the reference arm of the interferometer by
means of the piezoelectric actuator \textbf{PZT}. The recorded
intensities of the interferograms corresponding to the different phase
shifts, can be described as \cite{Creath1988}:
\begin{align}\label{eq:psi-one}
  I_{\ell}(x,y)  = &I_0(x,y) \left\lbrace \vphantom{\gamma(x,y) \cos\left[ \varphi(x,y) - \frac{\pi}{4} +  \frac{\pi}{2}\ell\right]} 1 + \right. \nonumber \\
                   &\left. \gamma(x,y) \cos\left[ \varphi(x,y) - \frac{\pi}{4}
                     +  \frac{\pi}{2}\ell\right]\right\rbrace , ~\ell=1,2,3
\end{align}

where $(x,y)$ represents the transverse position in the output plane
\ImagePlaneC, $I_0(x,y)$ is the arithmetic sum of the intensity of the
light beams in each arm of the interferometer, $\varphi(x,y)$ is their
relative phase and $\gamma(x,y)$ is the modulation of the interference
fringes. From these three interferograms it is possible to obtain the
relative phase of the object beam (\IA) with respect to the reference
beam (\FA), at every point of \ImagePlaneC
\begin{equation} \varphi(x,y) = \tan^{-1}\left(\frac{I_3(x,y)
      -I_2(x,y)}{I_1(x,y) - I_2(x,y)}\right). \label{eq:psi-phase}
\end{equation}
In our case, the phase over each slit region should be a
constant. However, there exist slight variations ( $\sim 2\%$) mainly due to
inhomogeneities of the LCoS display used as SLM, so we have taken as
argument of the coefficient $c_k$ in Eq.~(\ref{qudit}), the average of
the obtained phase, $\varphi_k = \overline{\varphi(x,y)}$, over the
interference region assigned to the slit $k$ (see
Fig.~\ref{fig:exp_photo_interf}). It should be considered that when
applying the PSI algorithm the recovered phase is not $\varphi_k$ but
$\varphi_k-\varphi_0$. Hence, for reconstructing the quantum state up
to a global phase, we can always define the phase of the reference
slit, $\varphi_0$, as zero. The modulus of the coefficients $c_{k}~'s$
correspond to the square root of the slit intensities and can be
obtained just by blocking the reference arm and averaging over the
same ROIs.

It is obvious that the slit selected as reference, for a given quantum
state, must have a non-null intensity value. It means that the
presented algorithm fails
when the quantum state to be determined has a null coefficient $c_0$.
To prevent such a case, a possibility is to first measure the
intensity of the $d$ slits and obtain the modulus of each coefficient,
$\vert c_{k} \vert$; then, the slit with the greater intensity value
can be selected as the reference and accordingly, adjusting the
position of \SFdos. The drawback of this strategy is that the
reference must be redefined every time, what entails to change the
filter position and realign the setup during the measurements.  In
order to avoid that, which is experimentally not convenient and
time-consuming, we adopted an alternative possibility which consists
in adding an extra slit with maximum transmission amplitude to be used
as reference, totalling $d+1$ slits of which only $d$ are used to
codify the state. Hence, we are able to reconstruct arbitrary pure
states without changing the experimental configuration.  Besides, with the
addition of these intensities measurements, we can distinguish between
pure and mixed states. Pure states are characterized by interference
patterns with maximum visibilities (bounded to the ratio of
intensities between interfering beams) and denote maximum coherence
between any pair of slits, whose value can be easily obtained from the set of measurements outcomes \cite{Creath1988}.

\section{Results and discussion}\label{results}

\begin{figure}
\centering
\includegraphics[width=\linewidth]{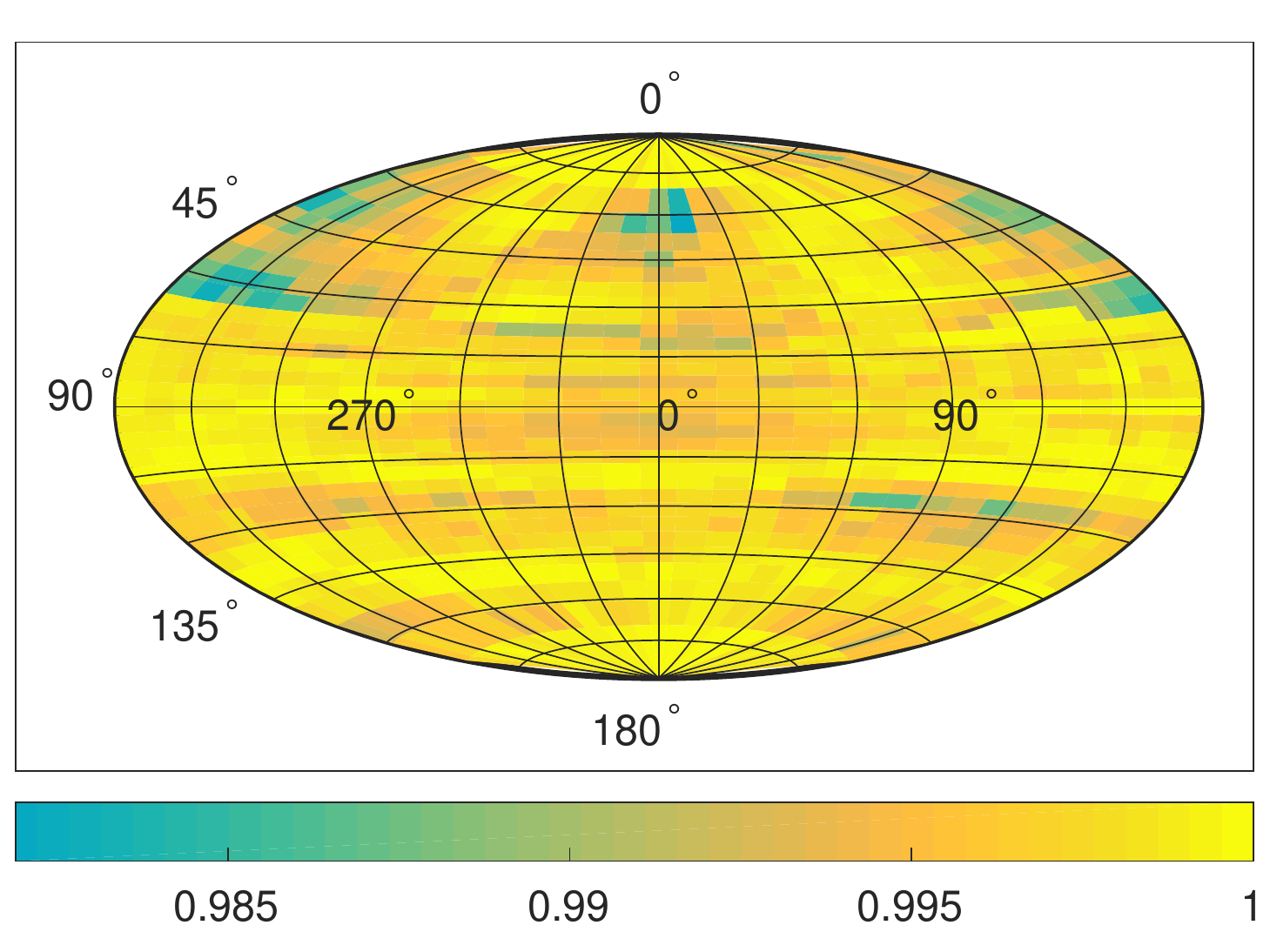}
\caption{Bloch sphere showing the reconstruction fidelities of 1024
  states uniformly distributed on the surface. The mean value of the
  fidelity is $\overline{F} = 0.997$, and the standard deviation
  $\sigma_F= 0.003$.\label{fig:bloch}}
\end{figure}

\begin{figure}
\centering
\includegraphics[width=\linewidth]{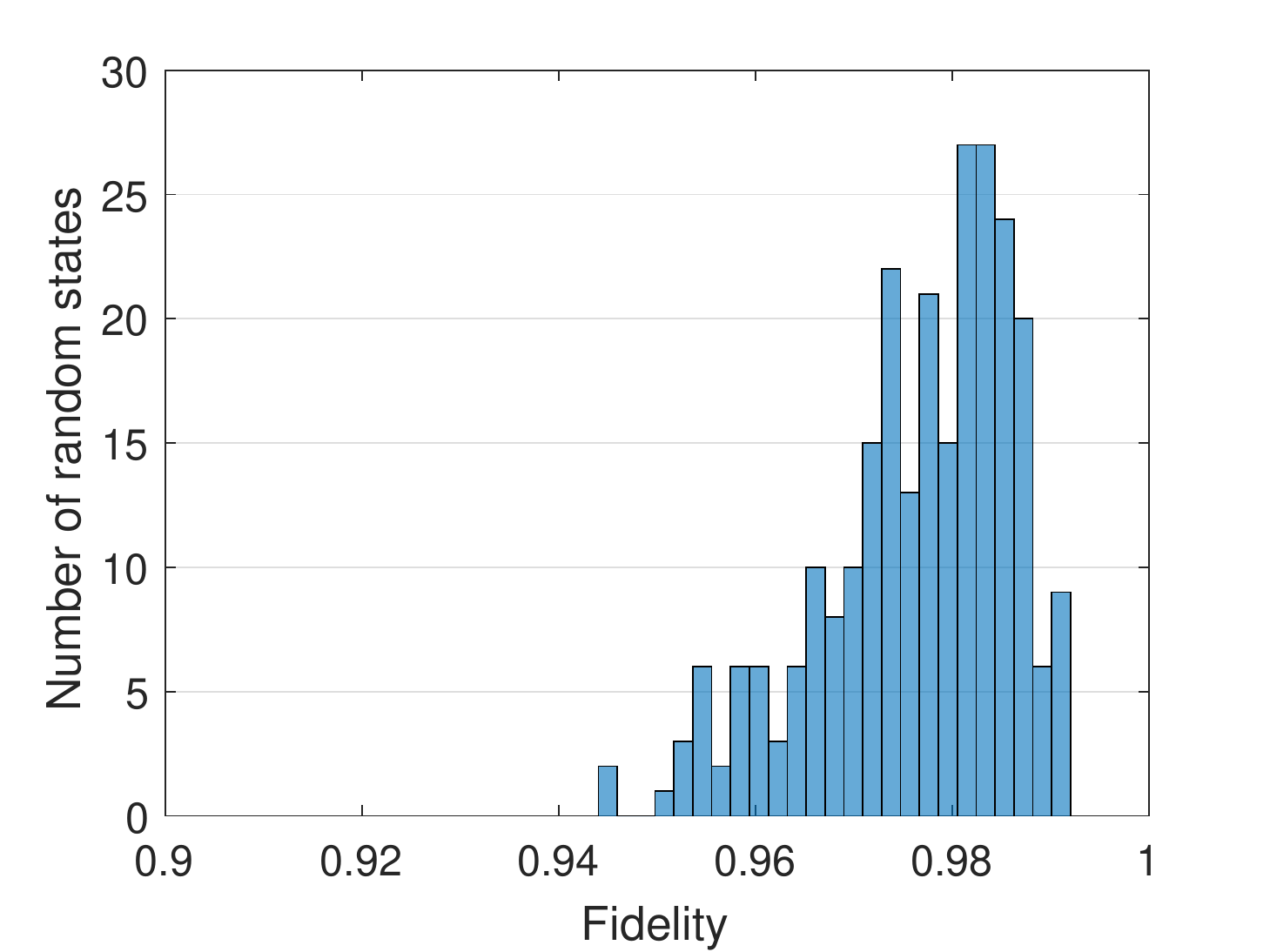}
\caption{Histogram of the reconstruction fidelities for 250 random states in d=14. The mean value of the fidelity is $\overline{F} = 0.98$, and the standard deviation $\sigma_F = 0.01$.\label{fig:histogram}}
\end{figure}

To evaluate the viability of the method and the quality of the
proposed setup we performed the reconstruction of a large number of
pure states, taking as examples systems of dimension $d=2$ and
$d=14$. As figure of merit, we calculated the fidelity
$F \equiv
\mathrm{Tr}\left(\sqrt{\sqrt{\varrho}\rho\sqrt{\varrho}}\right)$,
between the state intended to be prepared, $\varrho$, and the density matrix of the
reconstructed state, $\rho$ \cite{Jozsa1994}. Ideally, $F =1$. Figure
\ref{fig:bloch} represents the obtained fidelities for 1024 qubits
($d=2$) uniformly distributed in the surface of the Bloch sphere. The
mean value of the fidelity is $ \overline{F} = 0.997$, and the
standard deviation $\sigma_F = 0.003$. The histogram in
Fig. \ref{fig:histogram} shows the occurrence of the fidelities for
250 states of dimension $d=14$ randomly chosen. The average fidelity
is $ \overline{F} = 0.98$ while the standard deviation is
$\sigma_F = 0.01$. In this high-dimensional case the mean fidelity is
only slightly lower than in the bidimensional case. A similar
behaviour was observed for qudits of intermediate dimensions not shown
here. Then, the limitation of the experimental setup for implementing
the reconstruction method is the amount of slits that falls under the
central diffraction pattern of the reference slit. In order to verify
the purity of the states we have compared the actual visibility,
obtained from $\gamma(x,y)$, with the expected value for a pure state
that can be calculated from the intensity measurements. We have
observed that they overlap within the experimental errors.

It is worth to note the relation between the classical PSI steps and
quantum projectors. For every slit $k$ - which defines the state
$\ket{k}$ of the canonical base - except the reference, we can define
a set of three $d$-dimensional states
\begin{equation}
  |\Psi_{\ell}^{(k)}\rangle =\frac{\ket{0} + e^{i\pi/2\times\left(\ell-1/2 \right)} \ket{k}}{\sqrt{2}},~\ell=1,2,3
\end{equation}
where $\ket{0}$ represents the reference slit, and $k$ runs from 1 to
$d-1$. These states show the same phase relation between the reference
and the target slit that the phase shifts introduced in the three step
PSI. %and thus represent each of the three PSI steps.
To each of these states we can associate a projector
$\PP{\ell}{k}=|\Psi_{\ell}^{(k)}\rangle\langle\Psi_{\ell}^{(k)}\vert$. The
outcome probabilities of this set of projectors,
$p_{\ell}^{(k)} =
\bra{\Psi}\PP{\ell}{k}\ket{\Psi}=\vert\langle\Psi_{\ell}^{(k)}\ket{\Psi}\vert^2$,
are given by the following expression, totally analogue to those
described in Eq.~\eqref{eq:psi-one},
\begin{align}
p_{\ell}^{(k)} &=& \frac{|c_0|^2}{2} + \frac{|c_k|^2}{2} + \Re\lbrace c_0 c_k^* e^{i \pi/2\times\left(\ell-1/2 \right)}\rbrace.
\end{align}
With the knowledge of $c_0\equiv +\sqrt{p_0}>0$, which is obtained
from the probability $\vert\langle0\ket{\Psi}\vert^2=p_0$, any $c_k$
is determined by means of the expression:
\begin{align}
\sqrt{2}c_0c_{k}^*&=&\left(p_{1}^{(k)}-p_{2}^{(k)}\right)+i\left(p_{3}^{(k)}-p_{2}^{(k)}\right).
\end{align}
Thus, the measurement outcomes of these
$3(d-1)$ projectors in addition with a previous measurement onto
the canonical base ${\vert k \rangle}_{k=0}^{d}$, are enough to
determine any pure state and certify the \textit{a priori} assumption
of purity. As these projectors do not depend on the nature of the
quantum system the tomographic scheme is not restricted to the present
setup and it can be in principle implemented for general quantum
systems.

\section{Conclusion}

Summarizing, we have presented a method that reduces to a minimum the
number of measurements for reconstructing all pure quantum states of
arbitrary dimension $d$. For this tomographic scheme the outcome
probabilities of a total of $4d-3$ projectors are needed, from which we
can also certify if the quantum system is actually in a pure state. Moreover, in the particular case of spatial qudits we propose and
experimental implemented a setup that enables to perform this method
in a non adaptive way and reducing the number of measurement outcomes
to only four, independently of the dimension $d$ of the states to be
characterized. We have observed a quite good performance of our
implementation at least up to dimension $d$ = 14, with mean fidelities
between the expected and reconstructed states higher than 0.97 in any case.

\begin{acknowledgments}
We express our gratitude to C. T. Schmiegelow for providing us with the sCMOS camera. This work was supported by UBACyT 20020130100727BA, CONICET PIP 11220150100475CO, and ANPCYT PICT 2014/2432. QPS is in a CONICET Fellowship.
\end{acknowledgments}

\bibliography{qt-psi-paper}
%\bibliography{bibliography}

\end{document}